\documentclass[a4paper,11pt]{article}
\usepackage{aaskaiid}
\usepackage{graphicx} 
\usepackage{xcolor}
\newcommand{\msun}{\mbox{$M_\odot$}}
\newcommand{\hi}{\mbox{H\,{\sc{i}}\,}}
\newcommand{\hii}{\mbox{H\,{\sc{ii}}\,}}
\newcommand{\cmc}{\mbox{cm$^{-3}$\,}}
\usepackage{graphicx}
\usepackage{natbib}
\usepackage{aas_macros}

\usepackage{amssymb}
\usepackage{hyperref} 
\usepackage{floatrow}
\usepackage{orcidlink} 

\title{The Impact and Environment of Massive Stars and Stellar Clusters}
\ShortTitle{The Impact of Massive Stars}

\ShortName{Anderson et al.} 
\author[1,2,3]{Loren Anderson\, \orcidlink{0000-0001-8800-1793}}
\author[4]{Jagadheep D. Pandian\, \orcidlink{0000-0003-4031-1121}}
\author[5]{Jyotirmoy Dey\,\orcidlink{0000-0003-4074-4365}}
\author[6]{Marco~Padovani\,\orcidlink{0000-0003-2303-0096}}
\author[7]{Ramlal Unnikrishnan\,\orcidlink{0000-0002-2843-2476}}
\author[8,9]{Annie Zavagno\, \orcidlink{0000-0001-9509-7316}} 
\author[10]{Grazia Maria Umana\, \orcidlink{0000-0002-6972-8388}}
\author[11]{Jakob van den Eijnden\orcidlink{0000-0002-5686-0611}}
\author[6]{Giovanni Sabatini\orcidlink{0000-0002-6428-9806}}
\author[12]{Kimberly L. Emig\, \orcidlink{0000-0001-6527-6954}}
\author[13]{Alessio Traficante\, \orcidlink{0000-0003-1665-6402}}
\author[14,15]{\'Alvaro S\'anchez-Monge\orcidlink{0000-0002-3078-9482}}
\author[16,17]{D. Anish Roshi\, \orcidlink{0000-0002-1732-5990}}

\affiliation[1]{Department of Physics and Astronomy, West Virginia University, Morgantown, WV 26506, USA}
\emailAdd{loren.anderson@mail.wvu.edu}
\affiliation[2]{Center for Gravitational Waves and Cosmology, West Virginia University, Chestnut Ridge Research Building, Morgantown, WV 26505, USA}
\affiliation[3]{Adjunct Astronomer at the Green Bank Observatory, P.O. Box 2, Green Bank, WV 24944, USA}
\affiliation[4]{Department of Earth and Space Sciences, Indian Institute of Space Science and Technology, Trivandrum 695547, India}
\affiliation[5]{Department of Astronomy \& Astrophysics, Tata Institute of Fundamental Research, Mumbai, 400005, India}
\affiliation[6]{INAF-Osservatorio Astrofisico di Arcetri, Largo E. Fermi 5, 50125, Firenze, Italy}
\affiliation[7]{Department of Space, Earth and Environment, Chalmers University of Technology, 412 96, Gothenburg, Sweden}
\affiliation[8]{Aix Marseille Univ, CNRS, CNES, LAM Marseille, France}
\affiliation[9]{Institut Universitaire de France, 1 rue Descartes, 75005 Paris, France}
\affiliation[10]{INAF-Osservatorio Astrofisico di Catania, Via Santa Sofia 78, I-95123 Catania, Italy}
\affiliation[11]{Anton Pannekoek Institute for Astronomy, University of Amsterdam, Science Park 904, 1098 XH Amsterdam, The Netherlands}
\affiliation[12]{National Radio Astronomy Observatory, 520 Edgemont Road, Charlottesville, VA 22903, USA}
\affiliation[13]{INAF-IAPS, Via Fosso del Cavaliere, 100, 00133 Rome, Italy}
\affiliation[14]{Institut de Ci\`encies de l'Espai (ICE), CSIC, Campus UAB, Carrer de Can Magrans s/n, E-08193, Bellaterra, Barcelona, Spain}
\affiliation[15]{Institut d'Estudis Espacials de Catalunya (IEEC), E-08860, Castelldefels, Barcelona, Spain}
\affiliation[16]{Florida Space Institute, University of Central Florida, Orlando, Florida 32826, USA}
\affiliation[17]{Center for Advanced Research in Science and Engineering (CARSE), University of Puerto Rico, Mayag\"uez, P.R. 00681, USA}

\begin{document}

\abstract{
    Massive stars and stellar clusters shape galactic evolution through powerful feedback mechanisms including radiation pressure, photoionization, stellar winds, and cosmic ray acceleration. However, their impact remains poorly understood due to observational challenges: they are rare, distant on average, and deeply embedded within dense, dusty environments. Radio observations provide a unique window into these processes, as radio emission penetrates obscuring material and traces both thermal free-free emission from ionized gas and non-thermal synchrotron emission from shocks and particle acceleration.
The Square Kilometre Array (SKA) will revolutionize massive star studies through unprecedented sensitivity and angular resolution. SKA observations will enable detailed characterization of hierarchical structures within \hii\ regions, measurements of physical conditions through hydrogen, helium, and carbon radio recombination lines (RRLs), and detection of non-thermal emission from cosmic ray acceleration in star-forming regions. SKA will permit systematic measurements of stellar wind mass-loss rates, studies of photoionized gas kinematics and dynamics, and exploration of photodissociation regions surrounding ultracompact \hii\ regions. Additionally, magnetic field strengths can be probed through Zeeman effect observations of RRLs.
This chapter discusses the current understanding of massive stars and stellar clusters and their feedback processes. We highlight how SKA observations will advance our knowledge of massive star formation, stellar winds, hierarchical structures in \hii\ regions, cosmic ray acceleration, and magnetic field regulation of star formation—providing crucial insights into feedback mechanisms governing the structure and evolution of the Milky Way and galaxies.}

\maketitle

\section{Introduction} 

The impact of massive stars ($M \geq 8$ \msun) and stellar clusters is not well understood. Observing these objects is significantly more challenging than their low-mass counterparts. Massive stars form quickly and in dense, often chaotic environments \citep[e.g.][]{2003ApJ...585..850M,2007ARA&A..45..481Z,2018ARA&A..56...41M, Sabatini21}. To make matters more difficult, because they are rare, massive stars tend to be far away, pushing the limits of what current telescopes can resolve and detect. Gaining insight into the formation and impact of massive stars is essential, as they play a major role in shaping the evolution of galaxies through the emission of jets, the ejection of material, the generation of stellar winds, and the accretion of gas from the surroundings \citep{2005ApJ...618L..33K,2007ARA&A..45..481Z,2011BSRSL..80..211K,2016ARA&A..54..491B}.

While much progress has been made through observations and simulations, key questions remain regarding the earliest phases of massive star and cluster formation and their impact on their environments. The short lifetimes and deeply embedded nature of massive protostars make it challenging to observationally distinguish between formation scenarios such as core accretion and competitive accretion \citep{2003ApJ...585..850M,2006MNRAS.370..488B,2019ApJ...886..102S}. Feedback processes, including radiation pressure, ionized winds, and protostellar outflows, are expected to disrupt accretion, although evidence suggests that accretion can persist through dense, self-shielded structures \citep{2009Sci...323..754K,2018A&A...620A.182K}. Additionally, in clustered environments, feedback from massive stars can reshape their natal clouds, driving turbulence, or triggering secondary star formation \citep{2005A&A...433..565D,2006A&A...446..171Z,2012MNRAS.424..377D,2015A&A...573A.106G}. These open questions underscore the need for high-resolution, wide-field observations that can simultaneously trace the feedback mechanisms, such as ionized jets and winds from massive stars and stellar clusters, and their impact on the surrounding environment.

Radio observations provide an unparalleled means of investigating these processes. Unlike optical or infrared wavelengths, radio emission penetrates the dense, dusty environments in which massive stars form. Thermal free–free emission traces ionized gas within \hii regions, while non-thermal synchrotron emission reveals shocks and sites of particle acceleration. Radio recombination lines (RRLs) of hydrogen, helium, and carbon provide diagnostics of temperature, density, and kinematics, and can constrain turbulence and magnetic fields that influence the stability of molecular clouds and efficiency of feedback from massive stars.

The Square Kilometre Array (SKA) will enable an unprecedented advance in this field. The excellent sensitivity and angular resolution of the SKA allows for detailed studies in both continuum and spectroscopic observations. Such capabilities will permit the detailed characterization of ionized gas morphologies and kinematics, the detection of non-thermal emission associated with shocks and cosmic-ray acceleration, and the measurement of magnetic field strengths through Zeeman effect observations. As the topic of a Galactic plane continuum survey is addressed in the Chapter ``The 10-15 GHz Radio Continuum
Survey of the Galactic Plane with SKAO'' \citep{Traficante01.2026.SKA} and that of spectroscopic studies in the Chapter ``Spectroscopic Surveys with the SKA Probing the Ionized and Molecular Milky Way'' \citep{Karska01.2026.SKA}, here we focus on how such observations would allow for a detailed study of the impact massive stars and clusters have on their environments.

In the following sections, we examine the current understanding of massive stars and stellar clusters, their feedback processes and hierarchical structures in \hii regions (Section~\ref{msc-details}), and the role of magnetic fields and physical conditions near ionized–molecular interfaces (Section~\ref{bfield-hii}). Together, these discussions highlight the critical role of upcoming SKA observations in advancing our knowledge of massive star formation and feedback within the Galactic context.


\section{Massive stars and stellar clusters}
\label{msc-details}
Massive stars of spectral types O and B produce copious amounts of radiation.  Stars of spectral type earlier than B2 can fully ionize their surroundings, creating \hii regions. More than 8\,000 Galactic \hii regions are known \citep{2014ApJS..212....1A}. This number is likely an underestimate because the census is incomplete due to the very high sensitivity and spatial resolution required to detect the earliest phases.

At later stages of the evolution of massive stars, e.g., after their formation and particularly after moving away from the main sequence, their radio properties are dominated by their strong stellar winds. Various avenues exist through which these winds drive low-frequency emission: thermal emission from the radio photospheres of individual massive stars, shock-powered non-thermal emission from colliding winds in massive binaries, or different types of nebulae driven by the stellar winds interacting with the interstellar medium. The faint nature of such (extended) emission, particularly from the latter feedback structures, necessitates a combination of high sensitivity across a broad frequency range with an array design optimized for the detection and imaging of low-surface brightness structures\footnote{Here, we focus on the point-source emission the star itself. The alternative radio methods to understand stellar winds, via their interaction in binaries and with surrounding ISM, are discussed in depth in the Chapters ``Accreting Compact Object Binaries with the SKA'' \citep{Beri01.2026.SKA} and on ``Evolved Massive Stars and their Impact on their Environment'' \citep{Buemi01.2026.SKA}.}.

The radio emission from stellar winds of single/non-interacting massive stars is dominated by the thermal Bremsstrahlung originating from the outer regions of the wind. For a smooth wind that has reached its constant terminal velocity at a smaller radius than where the wind become optically thin to this emission, the density-squared Bremsstrahlung emissivity leads to an expected power law spectrum with slope $\alpha = 0.6$ \citep{1975MNRAS.170...41W,1975A&A....39....1P}. Deviations from this slope may arise due to the contributions of non-thermal processes with steep spectra, particularly below $\sim 1$ GHz \citep[leading to $\alpha < 0.6$;][]{2022ApJ...932...12E}. Alternatively, if the wind has not yet reached its terminal velocity but still accelerates in the low-frequency emission region, the spectral index will be $\alpha > 0.6$ \citep[as observed using, e.g., radio and sub-mm observations;][]{2025MNRAS.543..862V}. The overall brightness of the SED is set by the mass-loss rate and velocity profile, which together determine the wind's radial density profile. Deviations from a smooth wind, i.e., clumps, lead to incorrectly modeled normalization of the spectrum, as these affect the density structure. However, since clumping is expected to decrease with distance from the star, this effect is expected and observed to be of a smaller magnitude than for inner-wind diagnostics \citep[e.g.,][]{2022A&A...658A..61R}. 

\begin{figure}
\centering
\floatbox[{\capbeside\thisfloatsetup{capbesideposition={right,top},capbesidewidth=6.1cm}}]{figure}[\FBwidth]
{\caption{The minimum detectable stellar wind mass loss rate with SKA-mid in AA4. For all OB stars detected with Gaia within 1 kpc \citep{2025MNRAS.538.1367Q} we calculate what wind mass loss rate is detectable at Band 5a in 10 minutes (black line) and 1 minute (red line) of observing time, In blue, we show only hot stars ($T \gtrsim 15000$ K). We note that all histograms shift only slightly in Band 5b or when assuming AA$^{*}$ instead. Here, we assume relative fast winds ($1000$ km/s), while for slower winds, the minimum detectable mass loss rates becomes lower. Currently, mass loss rates are usually detected when of the order $10^{-5}$ $M_\odot$/yr. \label{fig:stellarwinds}}}
{\includegraphics[width=9cm]{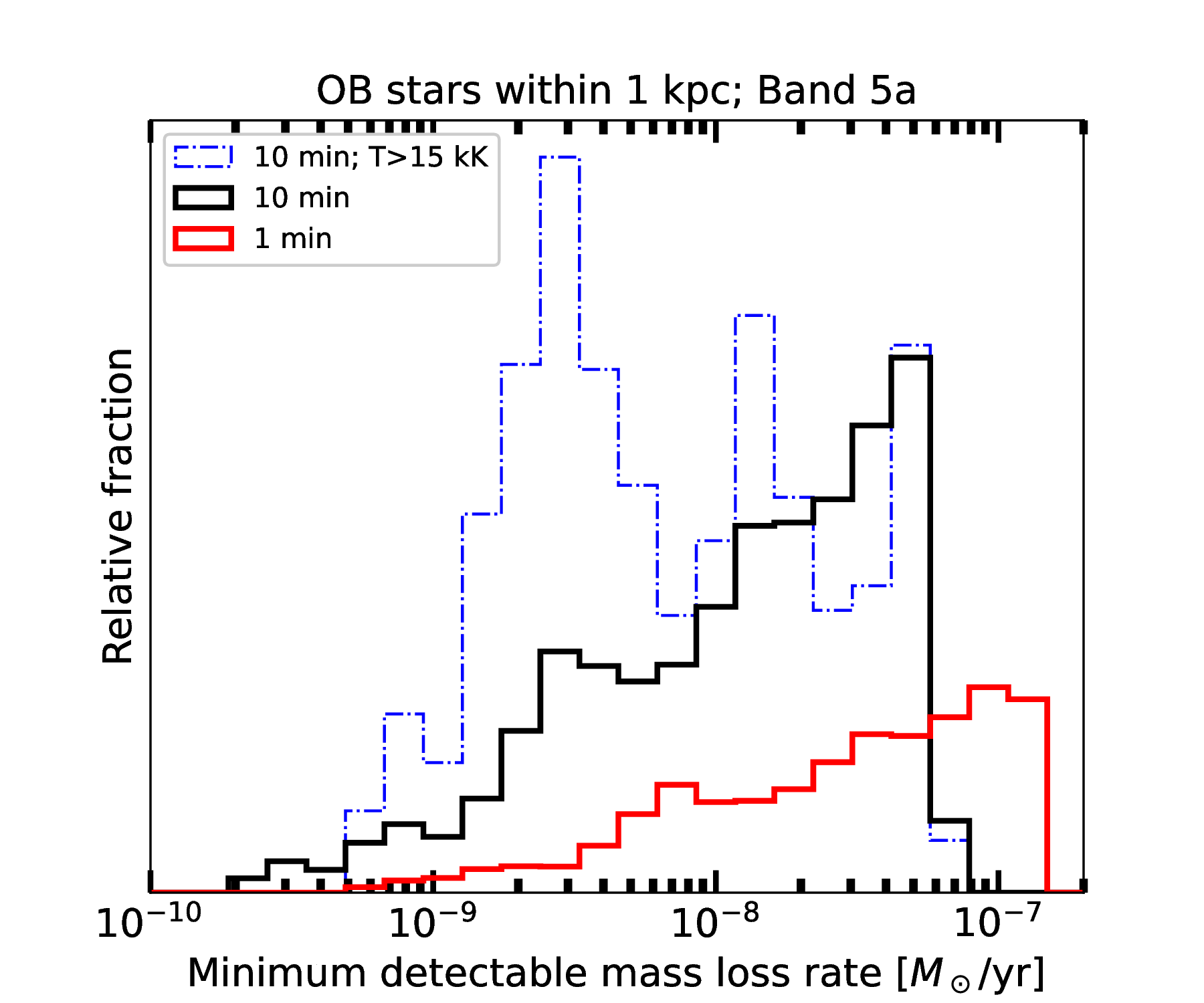}}
\end{figure}

Low-frequency radio observations are therefore a powerful tool for constraining stellar-wind mass-loss rates and velocity profiles. However, this approach remains relatively underexplored
\citep[e.g.,][]{2018A&A...617A.137F,2019A&A...632A..38A,2021A&A...647A.110G}. Due to the inverted nature of the emission, most studies are performed with a combination of radio ($\gtrsim 5$ GHz) and (sub-)mm facilities. SKA precursor telescopes observing around $\lesssim 1$ GHz have the advantage of a larger field of view, but lack the sensitivity for systematic detections of massive stars. The SKA, particularly SKA-mid, offers the required enhancement in sensitivity and range in observing frequency to systematically detect and characterize the stellar winds of nearby massive stars at radio frequencies. We demonstrate this SKA improvement in Figure \ref{fig:stellarwinds}, highlighting the typical expected stellar wind flux densities versus the expected sensitivities. 

\subsection{Feedback} 
\label{subsec:clusters_and_feedback}

Feedback associated with massive stars during their lifetimes includes radiation pressure, photo-ionized gas, and stellar winds. SKA observations will be particularly impactful for studying these phenomena through measurements of the radio continuum emission and radio recombination lines.  



Photoionizing radiation pressure is expected to have large contributions in the earliest stages of an \hii\ region's evolution and/or cloud material that is in very close proximity (on the order of $\sim$0.1 pc) to a massive star. At later times and/or when an \hii\ region has expanded, the gas densities are lower and porous/non-homogeneous material allows for ionizing radiation to escape the immediate surroundings and heatup and ionize the ISM of the Galaxy.

Photoionizing radiation from massive stars heats the cold ($T \sim 10$~K), dense surrounding material surrounding to temperatures of $T \sim 10^4$~K.  This increases the thermal pressure by orders of magnitude, which can mechanically unbind the gas. In more evolved regions and/or for blister-like regions at the edge of a molecular cloud, photoionized gas photoevaporates off the cloud, accelerating the flow of material. Quantifying the mass flow rate of this effect is important for understanding stellar feedback and its influence on galaxy evolution. In \hii regions with radii $\gtrsim$ a few pc, photoionized gas may dominate the pressure and dynamics of the region \citep{Olivier2021,Lopez2014ApJ}. Systematic and spatially resolved measurements of the ionized gas pressure in \hii regions in the Galaxy are sorely needed.

SKA observations of the free-free continuum emitted from ionized gas will illuminate the distribution and structure of photoionised gas, providing a view into the dense gas environment at the SKA-mid frequencies (unattenuated by dust) and into the low density material at SKA-low frequencies ($n_e \lesssim 100$~\cmc). The photoionizing flux can be estimated from the free-free radio continuum, given that the recombination time is relatively short such that an active source of ionizing radiation incident on the material is needed to maintain its ionization.

Similarly Hydrogen and Helium radio recombination lines (HRRLs and HeRRLs, respectively) trace the ionized gas and in turn the photo-ionizing radiation incident on it. While longer integration times are needed to reach comparable emission measures for RRLs with respect to the continuum, an analysis of the spectral energy distribution is needed for the continuum (to measure the spectral index and infer if the emission is a mix of or dominated by free-free emission and/or synchrotron) whereas the RRLs arise exclusively from ionized gas. RRLs also have the added benefit of measuring the gas kinematics, key to quantify flow rates of the ionized material and identifying kinematically-coherent clouds. 
Low-frequency HRRLs may provide the most effective probe of low-density ionized gas, enabling direct constraints on the electron density (and thus the characteristic size scale). Since density is typically the dominant source of uncertainty, such measurements would significantly improve estimates of ionized-gas masses, photoionized mass-flow rates, and ionized-gas pressures.

\subsection{Hierarchical structures in Galactic HII regions}
\hii regions emit prominently at radio and infrared wavelengths, with radio emission arising from thermal bremsstrahlung. The bremsstrahlung radiation from a classical Str\"{o}mgren sphere manifests a spectral index (defined through $S\nu \propto \nu^\alpha$) of +2 and $-$0.1 in the optical thick and thin regimes, respectively. The transition between these regimes occurs at a frequency that is proportional to the emission measure of the \hii region. However, it has been observed that the spectral index in hypercompact (HC) \hii regions ranges from 0 to +2 at frequencies from 5 to 22~GHz, with a typical value around +1 (e.g. \citealt{franco2000}, \citealt{beuther_ppv_2007} and references therein).


The observation of a spectral index that is different from +2 can be due to different reasons. Since \hii regions form from ionization of material around the massive star, which in turn forms from cores that have a density profile, one can expect there to be density gradients in the ionized gas. If the electron density in the \hii region has a power law of the form $n_e \propto r^{-\omega}$ and the medium is optically thick at all frequencies in the central region, then the observed spectral index is expected to be $\alpha = (2\omega - 3.1)/(\omega - 0.5)$ \citep{olnon1975}. Although the density profiles of the cores have values of $\omega$ ranging from 1 to 3 (e.g. \citealt{arquilla1985}, \citealt{gieser2023}), it was found that a density stratification as steep as $r^{-4}$ is required to explain the spectral index towards some sources \citep{franco2000}, which is nonphysical.

\begin{figure}
\centering
\includegraphics[width=0.9\linewidth]{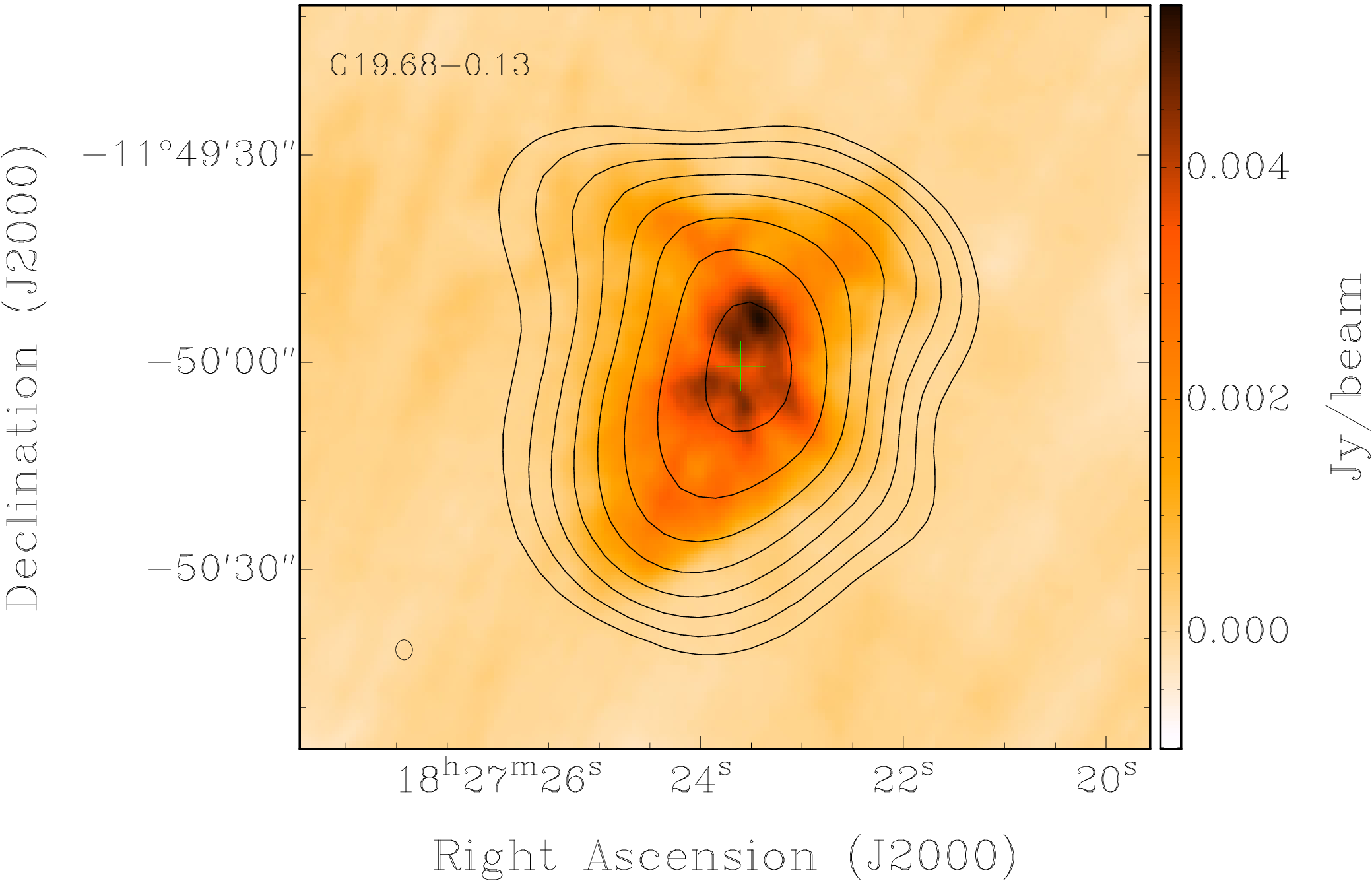}
\caption{The 1.35~GHz uGMRT radio continuum map ($\approx 2''$ resolution) of a \hii region, G19.68$-$0.13, overlaid with the 5.79~GHz radio continuum image from the GLOSTAR survey ($18''$ resolution) in black contours. The contours start at the 3$\sigma$-level and increase in steps of $\sqrt{2}$. The location of G19.68$-$0.13, as reported in the THOR radio continuum catalog, is indicated using a green `$+$' sign. The uGMRT beam size is shown at the bottom-left corner of the figure. Note the presence of radio continuum emission over a wide range of angular scales.}
\label{fig:radioemission}
\end{figure}

\begin{figure}
\centering
\includegraphics[width=0.75\linewidth]{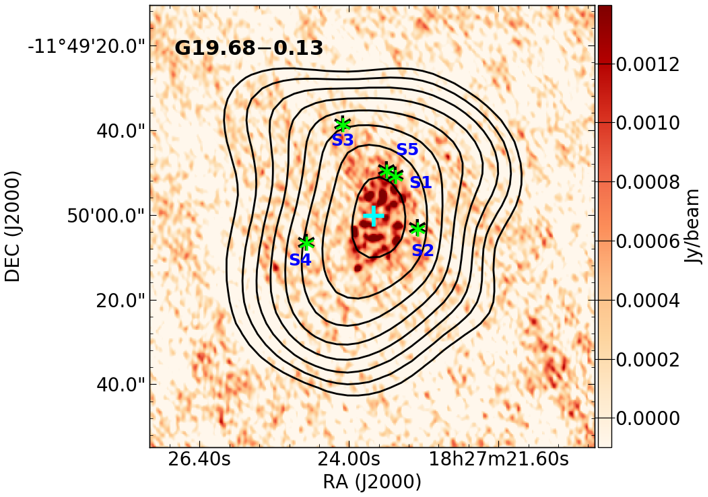}
\caption{The 5~GHz radio continuum emission from G19.68$-$0.13 as seen by the CORNISH survey with locations of the candidate ionizing stars overlaid. The candidate stars are identified using photometric data from 2MASS and UKIDSS surveys. Note that the candidate ionizing stars are not coincident with the compact radio continuum emission as would be expected if the former were the cause of the latter.}
\label{fig:stars}
\end{figure}

It is well known that the structure in dense molecular clouds is hierarchical comprising large-scale diffuse gas, dense clumps, and denser cores (e.g. \citealt{goldsmith2008}). However, it is not clear how the hierarchical structure is formed in the \hii regions since a classical expanding ionization front creates either a uniform density Str\"{o}mgren sphere or a region with a density gradient. A study of the dynamics of the ionized gas in a large sample of \hii regions that span a range of sizes is required to address this question.

Assuming that the compact component of the hierarchical structure has sizes and emission measures comparable to those of HC \hii regions (sizes $\sim 0.05$~pc, EM $\sim 10^9$~pc~cm$^{-6}$), observations of radio recombination lines with angular resolution better than 0.5$''$ are required to probe the dynamics of ionized gas. These observations need to be carried out in Band 5b of SKA since the continuum will be optically thick at lower frequencies. In order to estimate the observing time required with the SKA in its AA4 configuration, we use the following equations. The continuum optical depth is given by
\begin{equation}
    \tau_C = 8.24 \times 10^{-2} \left( \frac{T_e}{\mathrm{K}} \right)^{-1.35} \left( \frac{\nu}{\mathrm{GHz}} \right)^{-2.1} \left( \frac{EM}{\mathrm{pc~cm}^{-6}} \right)
    \label{eq_tauc}
\end{equation}
following the Altenhoff approximation \citep{altenhoff1960}. The peak optical depth of the radio recombination line is given by
\begin{equation}
    \tau_L = 1.92 \times 10^3 \left( \frac{T_e}{\mathrm{K}} \right)^{-5/2} \left( \frac{\Delta\nu}{\mathrm{kHz}} \right)^{-1} \left( \frac{EM}{\mathrm{pc~cm}^{-6}} \right)
    \label{eq_taul}
\end{equation}
following \citet{wilson2009}. The specific intensities of the radio continuum and recombination line (at its peak) are given by
\begin{eqnarray}
    I_C & = & B_\nu(T_e)(1 - e^{-\tau_C}) \label{eq_ic} \\
    I_L & = & B_\nu(T_e)e^{-\tau_C}(1 - e^{-\tau_L}) \label{eq_il}
\end{eqnarray}

Assuming an electron temperature of $10^4$~K and an emission measure of $10^9$~pc~cm$^{-6}$, eqs.~(\ref{eq_ic}) and (\ref{eq_il}) suggest that the recombination line will have a specific intensity of $\sim 8.5 \times 10^{-15}$~erg~s$^{-1}$~cm$^{-2}$~sr$^{-1}$~Hz$^{-1}$. A region of size 0.05~pc will have an angular size greater than $\sim 1''$ (taking distances $\lesssim 10$~kpc, which is significantly larger than the native resolution of the SKA in its AA4 configuration at 12~GHz ($\sim 0.08'' \times 0.07''$). The correlator allows for a continuum bandwidth of 5~GHz and a spectral resolution of 13.44~kHz (0.34~km~s$^{-1}$). Since recombination lines have linewidths greater than 30~km~s$^{-1}$, one can adopt a spectral smoothing of 8 channels (which yields a velocity resolution of 2.7~km~s$^{-1}$) and target 11 recombination lines within the 5~GHz bandwidth (e.g. H76$\alpha$ to H86$\alpha$). This allows one to obtain a boost in signal-to-noise ratio of at least 3 by stacking the different recombination lines.

Since the targeted compact emission regions are expected to have an angular size $\gtrsim 0.5 ''$, the data can be smoothed to an angular resolution of 0.2$''$ (or higher for more nearby regions) and still achieve the goal of studying the dynamics of ionized gas. At a resolution of $0.2''$, a specific intensity of $\sim 8.5 \times 10^{-15}$~erg~s$^{-1}$~cm$^{-2}$~sr$^{-1}$~Hz$^{-1}$ is equivalent to 900~$\mu$Jy~beam$^{-1}$. Taking into account the boost in the signal-to-noise ratio by smoothing the different recombination lines, one can achieve a $10\sigma$ detection in an integration time of $\sim 0.5$~hr. Smoothing to a coarser angular resolution allows one to achieve either a better signal-to-noise ratio, or the same signal-to-noise ratio in a smaller integration time.

Although there is expected to be a range of emission measures associated with the hierarchical structure in \hii regions, the diffuse emission is expected to have an emission measure $\lesssim 10^6$~pc~cm$^{-6}$ since the emission is observed to be optically thin at 1~GHz. Radio recombination lines are also seen to have lower linewidths of 25$-$30~km~s$^{-1}$. One can utilize Band 2 of the SKA with correlator in continuum mode, since this provides a velocity resolution of around 3~km~s$^{-1}$ at the central frequency of 1.355~GHz, which is adequate to resolve the lines. The full continuum bandwidth of 0.81~GHz covers 35 radio recombination lines (H155$\alpha$ to H189$\alpha$), the stacking of which is expected to provide an improvement in signal-to-noise ratio of at least 5, accounting for the loss of some lines due to RFI. Utilizing eqs.~(\ref{eq_tauc}) to (\ref{eq_il}), the specific intensity of the recombination lines is found to be greater than $\sim 5 \times 10^{-18}$~~erg~s$^{-1}$~cm$^{-2}$~sr$^{-1}$~Hz$^{-1}$, or 1.2~mJy~beam$^{-1}$ at an angular resolution of 10$''$. Taking into account the boost in the signal-to-noise ratio by stacking multiple lines, this can be detected at a peak signal-to-noise ratio of 10 in an integration time of 20~minutes. 

The above sensitivity calculations demonstrate the ability of the SKA to carry out a systematic study of a relatively large sample of \hii regions to study the hierarchical structures within. Existing facilities do not have the sensitivity to investigate the velocity field of the ionized gas at the angular resolution above. A conclusive study of hierarchical structures also requires a synergy with millimeter/sub-millimeter facilities such as ALMA and SMA in order to understand the interaction between the ionized gas and the clumpy molecular medium within which the massive stars form. When coupled with complementary \hi data, one can determine the confinement of \hii regions and the link between \hii regions and the warm ionized medium (WIM) of the Milky Way. One can also use this study to investigate the contribution of other anomalous emission mechanisms such as spinning dust (e.g. \citealt{dickinson2009, planck2014, paladini2015}) to the overall spectral energy distribution of \hii regions since such mechanisms will not contribute to RRL emission.


\subsection{The role of massive stars in accelerating cosmic rays}

Cosmic rays (CRs) are highly energetic charged particles that permeate the Galaxy. The origin of CRs remains mysterious, especially for those with energy above a few Petaelectronvolts (PeVs).  Tracing the birthplace of CRs is difficult because of their diffusive motion through the interstellar medium (ISM).  They are thought to originate from supernova remnants (SNRs), pulsars (PSRs) and their nebulae (PWNe) \citep{blasi13}, and binary systems, especially those containing compact objects (aka microquasars) \citep{alfaro24}.  However, none of these sources seem to be able to accelerate protons up to PeV energies. 

There is growing observational evidence for the presence of non-thermal emission in star-forming regions. In particular, the presence of a negative spectral index in protostellar jet shocks \citep{Ainsworth+2014,Rodriguez-Kamenetzky+2016,Rodriguez-Kamenetzky+2017,Sanna+2019} has sparked considerable interest in understanding the origin of this non-thermal emission \citep{Padovani+2015}.

The synchrotron emissivity expected from Galactic CR electrons is not sufficient to explain the observed flux densities. However, if a small fraction of gravitational luminosity can be used to accelerate cosmic rays locally, the latter can easily dominate over the interstellar cosmic ray flux \citep{Padovani+2016}. 
There are at least three main acceleration mechanisms to accelerate particles \citep{Marcowith+2016}: stochastic Fermi acceleration, first-order Fermi acceleration (also known as Diffusive Shock Acceleration, DSA) and magnetic reconnection. 
In particular, DSA has currently been identified as the likely main mechanism for accelerating thermal particles in protostellar jets \citep{Padovani+2020}. 
According to DSA, thermal particles are trapped between the magnetic field irregularities on either side of a shock. At each shock crossing, particles interact via head-on collisions, gaining energy systematically \citep{Drury+1996}. 

A similar theoretical approach has been adopted to explain the non-thermal emission observed in \hii regions. While the latter are commonly characterised by thermal emission 
\citep[e.g.][]{WoodChurchwell1989,Kurtz2005,SanchezMonge+2008,SanchezMonge+2011,Hoare+2012,Purcell+2013,Wang+2018,Yang+2019}, instruments such as VLA and GMRT discovered the presence of non-thermal emission in several such objects. This non-thermal emission typically occur in distinct spots embedded in thermal emission \cite[e.g.][]{Nandakumar+2016,Veena+2016} or as areas contigous to it \cite[e.g.][]{Mucke+2002}, as in the case of cometary \hii regions.

A special case is represented by the \hii region Sgr B2(DS), close to the Galactic centre, where \cite{Meng+2019} observed a clear synchrotron arc, without thermal contamination. This was interpreted as an expanding shock, at the front of which thermal electrons are accelerated up to relativistic energies. 
The results of this modeling are helpful in constraining physical quantities such as volume density, magnetic field intensity, ionisation fraction, flow velocity in the shock reference frame, and temperature \citep{Padovani+2019}.

The Galactic Centre region has recently been revealed as a particularly rich 
laboratory for this kind of study. The MeerKAT 1.28\,GHz survey of the inner 
Galaxy uncovered a vast population of non-thermal radio filaments, 
some of which are spatially associated with the boundaries of \hii regions and 
wind-blown bubbles driven by OB associations and young massive clusters 
\citep{Heywood+2022,YusefZadeh+2022}. The spectral indices of these filaments 
are consistent with synchrotron emission from GeV electrons.. 
Furthermore, MeerKAT revealed two large-scale radio lobes extending $\sim$430\,pc 
above and below the plane \citep{Heywood+2019}, whose edges coincide with enhanced 
non-thermal emission, suggesting that collective stellar feedback can inject CRs 
on scales far exceeding individual \hii regions.

The SKA will provide new constraints on CR acceleration in 
massive-star environments. It will deliver sub-arcsecond imaging with 
sufficient surface brightness sensitivity to resolve spectral index variations 
across the rims of \hii regions and non-thermal filaments, 
identifying shock acceleration sites through spectral index hardening 
\citep{Padovani+2019,Heywood+2022}. SKA will also 
complete the census of thermal and non-thermal shells in 
the Galactic Centre, allowing a statistical assessment of the relative CR 
injection rates from wind bubbles and \hii regions versus classical SNRs.

A few known TeV $\gamma$-ray sources have been found to be spatially coincident with clusters of young ($<10$\,Myr) stars, like Cygnus OB2, Westerlund~1 and Westerlund~2 \citep{aharonian19}. New theoretical efforts \citep{bykov20, vieu21, morlino21} and observations \citep{abesekara20, abeysekara21,aharonian22} have shown that acceleration inside star clusters (SCs) can reach ultra-high energies. Massive stars are characterized by severe mass loss through powerful stellar winds. The collective contribution of the wind of hundreds of stars in a SC excavates a bubble in the ISM generating a shock that endures for the entire life of the cluster. Depending on the number and type of stars, the stellar density and the medium in which the wind expands, favorable conditions for accelerating high energetic particles up to PeV energies can be found in SCs.  There may be as many as 100 $\gamma$-ray emitting star clusters in the Milky Way.

It is clear that the different values imply different sources contribute to the pool of GCRs. The main limitation in the estimation of acceleration efficiency, beyond the intrinsic uncertainties on the target gas and magnetic field, stands with the wind luminosity, that could only be constrained based on the theory of stellar evolution in the few objects for which the mass is known. This requires knowing the stellar population, which is well constrained only for evolved systems, where the extinction of light is limited \citep[e.g.,][]{celli24}. Even then, the limitations in the knowledge of the stellar mass distribution inside YMSCs limit the accuracy of the estimation \citep{krumholz19}. For other SCs, especially the youngest, which are at the same time also the most powerful in terms of winds, this information is often hidden beyond the layers of hot dust and gas. An extensive study of SC systems at different ages has not yet been undertaken.

To date, a number of low-frequency ($\lesssim10$~GHz) radio continuum surveys have been conducted. For example, the GLOSTAR survey \citep{Yang2023} confirmed that a substantial number of compact and extended regions throughout the Galaxy exhibit non-thermal emission. 
Thanks to its sensitivity and the capability to observe at even lower frequencies, SKA turns out to be the key facility to characterize non-thermal processes in greater depth. As discussed in Section~\ref{subsec:clusters_and_feedback}, such studies can provide estimates of stellar wind mass-loss rates, thereby yielding the wind luminosity required to constrain the efficiency of GCR acceleration.

\label{proposed_obs_stars}

\section{The physical conditions near HII regions}
\label{bfield-hii}
\subsection{H and HeRRL observations}
HRRLs and HeRRLs illuminate ionized gas and its kinematics. They are also very useful tools to measure the gas physical properties, which lie at the heart of forming a more complete model of processes associated with ionized gas and their role in the structure and evolution of our Galaxy. At SKA-mid frequencies, a measurement of the line-to-continuum flux density ratio measures the temperature of ionized gas. The temperature distribution in our Galaxy reveals the history of its chemical enrichment, as ionized gas temperatures are strongly dependent on the metallicity of the gas -- more metals, more cooling and lower temperatures. The line widths of HRRLs and HeRRLs are broadened by thermal and turbulent contributions; with a measure of the thermal component, the turbulent component can then be inferred, though it has yet to be done throughout the galaxy and in a spatially resolved manner.

From particularly dense gas, the RRLs from ionized gas at SKA-mid frequencies could also be pressure broadened, due to collisions among the electrons. This effect results in a Lorentzian rather than Gaussian profile, and additionally has a tell-tale frequency-dependent broadening of the line. This effect may be used to directly estimate the density of the emitting gas.

Non-LTE processes can affect RRL emission at ever decreasing frequencies (e.g., particularly $<$ a few GHz in the Milky Way). When the spectral line energy distribution (SLED) of low-frequency RRLs is measured across radio bands, the physical conditions (density, temperature, size) of the gas can be determined. This provides another avenue to determine the elusive densities of this gas phase, which are needed to construct a global understanding and model of ionized gas in the Milky Way.

\subsection{Carbon RRLs as probes of physical conditions and magnetic fields near UC HII regions.}

As mentioned in Section~\ref{subsec:clusters_and_feedback}, hydrogen and helium RRLs probe the ionized gas within the UC\hii regions. These regions expand due to their high internal gas pressure; however, their observed lifetimes exceed theoretical predictions of dynamical lifetimes by approximately two orders of magnitude. One plausible explanation for this discrepancy is the pressure confinement of UC\,\hii\ regions by surrounding warm, high-density molecular gas, which can provide sufficient external pressure to stabilize and prolong their evolution \citep[see][and references therein]{Roshi_et_al_2005}. Such pressure confinement would lead to several observable effects, including the heating of the surrounding neutral gas, leading to the formation of photodissociation regions (PDRs) around the UC\,\hii\ regions. Studying the physical properties of these regions is vital for understanding the early life and environments of massive stars. 

PDRs have been studied using dense molecular tracers, although the estimation of physical properties from such studies is limited by the high optical depth in the lines involved. However, the high electron density and warm temperatures of these regions make them ideal sites for the formation of carbon recombination lines (CRRLs), which arise from high principal quantum number transitions of singly ionized carbon (C$^{+}$) gas. CRRLs do not present high optical depth and hence provide an extinction-free tracer of the material in the PDRs, and provide an additional test to the theory that UC\,\hii\ regions are pressure confined. Further, it can facilitate detailed MHD models, explore the relationships between PDR formation and UC\,\hii\ morphologies, and greatly improve PDR modeling \citep[e.g.][]{Salas_et_al_2019}, overall contributing to a comprehensive investigation of high-mass star-forming regions.

Several single-dish and interferometric studies of CRRLs from PDRs in massive star-forming regions have been undertaken with telescopes including Arecibo and the VLA \citep[e.g.][]{Roshi_et_al_2005, Roshi_et_al_2007}. These efforts have been constrained by sensitivity limitations, as CRRLs are intrinsically weak. The SKA will significantly enhance these investigations by offering multi-frequency, spatially-resolved continuum measurements and simultaneous observations of H, He, and C RRLs towards UC\,\hii\ regions. UC\,\hii\ regions become optically thick below a few GHz, allowing accurate temperature estimates of the ionized gas from continuum observations. Combining this with models of hydrogen RRL formation, we can determine the emission measure and density of UC\,\hii\ regions. CRRLs are amplified by stimulated emission due to thermal background from UC\,\hii\ regions. Precise characterization of this background emission will allow modeling of the CRRLs, constraining the PDR’s temperature, density, and line-of-sight path length. To date, such modeling has only been achieved for a limited number of sources \citep[e.g.][]{Jeyakumar_and_Roshi_2013}. The SKA, with its enhanced sensitivity and Southern sky access, will extend this capability to a much broader range of UC\,\hii\ regions.

Magnetic fields are known to play a fundamental role in regulating the dynamics of molecular clouds. Observations of CRRLs toward PDRs have revealed non-thermal broadening in the line profiles \citep[see][]{Roshi_et_al_2007}, with linewidths exceeding those predicted by non-LTE thermal models by a factor of approximately two. This excess is hypothesized to result from the presence of MHD waves in the medium. If confirmed, CRRLs could provide a novel diagnostic for estimating magnetic field strengths in PDRs, as suggested by preliminary results from GBT and VLA observations of selected UC\,\hii\ regions \citep{Roshi_et_al_2007}. 

Multi-frequency CRRL observations with SKA-mid could validate this approach and enable comparisons with SKA magnetic field estimates derived from independent HI and OH Zeeman-effect measurements. Direct comparison with CRRL Zeeman observations could be attempted with the AA4 configuration and appears feasible for a few sources. To estimate the detection sensitivity, we used the predicted magnetic fields toward the sources listed in \citet{Roshi_et_al_2007} and their CRRL flux densities. We followed the method outlined by \citet{Troland1982} to estimate the sensitivity. Most UC\,\hii\ sources in the list exhibit CRRLs near 9 GHz. Using band 5b, smoothing the spectral resolution to 1/6$^{th}$ of the CRRL linewidth, averaging 12 RRLs within the receiver bandwidth, and integrating for 10 hours, the Zeeman signature from two sources (G32.80+0.19, G70.29+1.60b) could be detected at more than the 3$\sigma$ level. A UV taper was applied to obtain an angular resolution of $\sim$25$^{"}$ for the calculation. Zeeman splitting of CRRLs from more evolved PDRs, which produce brighter CRRLs at lower frequencies, appears to be even more promising with the AA4 configuration, both because of the brighter lines and narrower line widths. For example, the AA4 configuration in band 2 (averaging 8 RRLs, angular resolution $\sim$ 5$^{"}$ , 10 h integration) could image CRRL Zeeman splitting toward the source W3A. In the longer term, the enhanced sensitivity of SKA Phase 2 will permit direct Zeeman measurements of CRRLs toward a larger number of PDRs, allowing simultaneous and independent determinations of magnetic field strengths in these regions.



\section*{Summary}

This Chapter explored the impact that massive stars and stellar clusters have  on the structure and evolution of galaxies, and how the SKA will revolutionize studies of massive star feedback.  Radio observations uniquely penetrate the dense, dusty regions where massive stars form and evolve. They trace both the thermal free–free emission from ionized gas and the non-thermal synchrotron radiation from shocks and cosmic ray acceleration. SKA will enable systematic, high-fidelity studies of stellar winds, hierarchical structures in \hii\ regions, magnetic fields, and ionized gas kinematics across the Galaxy.  Together, these advances will provide a comprehensive view of how massive stars drive feedback and regulate the evolution of galaxies.

\bibliographystyle{abbrvnat-maxbibnames4}
\bibliography{reference}

\end{document}